\documentclass{article}
\usepackage{ijcai25}

\usepackage{times}
\usepackage{soul}
\usepackage{url}
\usepackage[hidelinks]{hyperref}
\usepackage[utf8]{inputenc}
\usepackage[small]{caption}
\usepackage{graphicx}
\usepackage{amsmath}
\usepackage{amsthm}
\usepackage{booktabs}
\usepackage[switch]{lineno}

\usepackage{amsfonts}

\usepackage[noend]{algpseudocode}
\usepackage{mathpartir}

\usepackage{xspace}
\def\sysname{{\textsc{SymLLM}\xspace}}

\newtheorem{example}{Example}

\newtheorem{proposition}{Proposition}
\newtheorem{corollary}{Corollary}

\title{LLM-Guided Compositional Program Synthesis}

\author{
    Ruhma Khan$^1$
\and
Sumit Gulwani$^2$\and
Vu Le$^{2}$ \and
Arjun Radhakrishna$^{2}$ \and
Ashish Tiwari$^{2}$ \And
Gust Verbruggen$^{2}$\\
\affiliations
$^1$Georgia Tech\\
$^2$Microsoft  Corp\\
\emails
mehek.ruhma@gmail.com,
\{sumitg, levu, arradha, astiwar, gverbruggen\}@microsoft.com
}

\newcommand\ignore[1]{{}}

\begin{document}

\maketitle

\begin{abstract}
 Program synthesis from input-output examples, also called programming by example (PBE), has had tremendous impact on automating end-user tasks. Large language models (LLMs) have the ability to solve PBE tasks by generating code in different target languages, but they can fail unpredictably. To recover for failure, most approaches, such as self reflection, use the LLM to solve the same task, but with a richer context.  We introduce a novel technique that recovers from failure by constructing simpler subtasks for the LLM to solve. Our approach performs compositional program synthesis using LLMs, where LLM not only guides the decomposition of the PBE task into subtasks, but also solves the subtasks.  We present different strategies for decomposing the original task. We experimentally show that our approach can solve about 30\% of the challenging task instances that are not solved by self reflection alone.  
\end{abstract}

\section{Introduction}

\ignore{
Programs are structured sequences of text with precise semantics.
They can be attached with formal statements of (full or partial) correctness,
which can potentially be automatically verified.
Generating programs is viewed as a challenging task as it is 
important to get both the structure and the semantics absolutely correct.
This is unlike what happens with natural language text, where 
minor mutations in a text still preserve its meaning and usefulness
to a large degree.
Consequently, generating programs is a widely popular research topic,
and it has been studied in-depth both in the Programming Languages
community, as well as the Machine Learning community, among others.
\endignore}

Program synthesis has a long history.
Initial work on program synthesis
focused on generating programs from formal 
logical specifications~\cite{manna1980deductive,sketch,cegis}.
Interest in program synthesis was renewed by the introduction of
programming by example (PBE), where the goal was to
generate programs from input-output specifications~\cite{gulwani2011automating}. 
The popular symbolic technique here was syntax-guided synthesis (SyGuS)~\cite{SyGus}. 
In  SyGuS, programs are enumerated following the syntax of the program
(provided in the form of a domain-specific language (DSL)). There are two ways
to enumerate programs: forward, also called bottom-up synthesis, starts from the inputs
and enumerates values that can be computed over the inputs~\cite{enumerative-deductive,eusolver} and 
backward, also called top-down synthesis, starts from the outputs and enumerates values that
could be used to compute the output~\cite{gulwani2011automating}.
Such symbolic approaches are restricted to synthesizing programs expressible in the fixed DSL.
They also have difficulty scaling to synthesis of large programs.

With the availability of pretrained large language models (LLMs), program synthesis 
is being revisited starting from natural language (NL) specifications, 
input-output tests, and a combination of the two~\cite{mbpp,humaneval}.
Neural approaches can address some of the challenges of symbolic methods listed above. 
%
\ignore{
There is a lot of benefit to using LLMs for
generating programs. First, LLMs are able to exploit knowledge from existing code
in public repositories, whereas it is very difficult to build symbolic procedures that
can use all that information. Second, LLMs provide very general solutions; for example, 
they can generate programs in different target languages, they can generate programs
for any class of applications, and so on. Finally, LLM-based solutions can
be developed cheaply -- there is no need to write and debug thousands of line of code.
\endignore}
In particular, LLMs are not restricted to synthesizing in any (domain-)specific language,
nor they face any scalability challenges.
Moreover, LLMs exhibit an emergent behavior, called few-shot learning, that gives it the
ability to perform a task from a few demonstrations (examples)
of the task. 
Due to these benefits, our goal here is to explore LLM-based techniques for classic
programming by example (PBE) problems.

\ignore{
While LLMs offer a lot of benefit, LLM-based solutions 
do not get even close to the precision of symbolic techniques.  
LLMs make mistakes. LLMs are also sensitive to prompts.
When one observes poor performance of an LLM-based approach on a PBE task
(or any task), it is often tempting to question the ``prompt'' and ask whether
the ``best'' prompt was used. In fact, there is a lot of work going on in
exploring prompt optimization and finding the ideal prompt for various tasks.
Based on our extensive experience working with LLMs, it is evident that
no prompt is perfect and there are often many good choices. It is 
extremely unlikely that there exists a golden prompt that remains consistently
superior to all other prompts for a class of tasks. Rather than tuning prompts, 
one needs to develop approaches that will make LLM-based solutions
more robust. In this paper, we develop an approach to make LLM-based program
synthesis more robust using inspirations from symbolic PBE techniques.
\endignore}

\ignore{
Before jumping to the solution, it is prudent to search for reasons why 
symbolic synthesizers are more reliable than LLMs. 
One plausible intuitive explanation
is that LLMs are not taught to ``look-ahead''
or ``backtrack'', which are two essential skills of a symbolic reasoning engine. 
For example, when generating the fifth line of code, an LLM does not go back and fix the first line, if need be.
To improve robustness of an LLM-based program synthesizer, it seems essential to add the ability to 
backtrack, when needed, and re-evaluate, re-think and repair midway through code generation.
We describe an approach that fills that gap to some extent.

\endignore}

\ignore{
In the vast space of symbolic techniques, we focus on a subclass of 
program synthesis domain called programming by example (PBE).
In PBE, we are given a set of input-output (IO) examples
and the goal is to synthesize a program
that can transform all inputs to the corresponding outputs.
We may be optionally given an additional set of inputs too, in
which case the generated program is also expected to work 
(not throw runtime exceptions) on the additional inputs and 
return non-null values. 
The key benefit from focusing on PBE is that we have a verifiable specification -- 
once any PBE approach returns a program, we can check if that program 
meets the input-output specification. 
Availability of a verifiable specification is a critical aspect of PBE.
For example, it enables a generate-and-check approach
where we enumerate programs and return one that meets the
specification. It also enables counter-example-guided synthesis, where if a 
program does not meet the specification, we use the failure to refine,
as in counter-example-guided inductive synthesis (CEGIS).
These benefits of PBE carry over from symbolic to LLM-based approaches:
LLMs can generate programs that can be filtered using the IO specification, and 
LLMs can also refine based on counter examples.
In fact, we will use such LLM-based systems as the baseline in this work.
The question we want to answer here is can we go beyond and improve
LLM-based approaches by exploiting our understanding of the workings of
symbolic approaches for PBE.

\endignore}

While LLMs offer a lot of promise, LLMs can fail unpredictably.
There is a lot of recent work on handling failures, such as
retrieval-augmented generation, self-reflection, and agentic architectures~\cite{rag,reflexion,autogen}.
However, all the methods, with the possible exception of the expensive agentic methods, are based on 
having the LLM retry solving the {\em{same task}}, but with some additional context.
We instead explore a different mechanism in this paper, namely decomposition of the problem.
How do we decompose the program synthesis problem?
The first observe that when LLMs fail, the failed
program still has pieces that could be reused~\cite{multimodal-regex-gpt3-oopsla21}.
In fact, since the program is generated from input-output examples, often the prefix
of the program (that processes the input) or the suffix of the program (that is
responsible for generating the output) are ``correct'' and can be salvaged.
Sometimes even the whole program is almost correct and can be salvaged.
We next observe that salvaging the prefix (or suffix) of a program maps nicely to performing
forward (or backward) synthesis, and thus we can use ideas from there.

We present our approach, \sysname, where when we get a wrong program from the LLM, 
we keep either the ``prefix'' or ``suffix'' of the wrong program and 
resynthesize the other part. Rather than use actual prefix or suffix of the code string, 
we actually parse the program and use the 
dataflow graph to find the first computations on the input leaves (which corresponds to the prefix of the program) and 
the last computations that produce the output (which corresponds to the suffix of the program).
Having extracted the salvaged pieces of the program,
our approach follows the play book of forward and backward synthesis to
compute the specification for the new PBE subproblem(s) using the  
execution semantics (forward propagation) of the prefix or
the inverse execution semantics (backward propagation) of the suffix.
Apart from these two strategies, \sysname also has two more strategies to decompose the synthesis task
and thus perform compositional program synthesis.
Our main finding is that compositional program synthesis can solve PBE tasks that are not
solvable by few-shot prompting and a self-reflection loop. 

\ignore{
Finally, to fully appreciate the power \sysname\ provides to LLMs, one needs to note that
both LLMs and symbolic PBE solvers follow the same principle: decompose the (program) generation task into subtasks
and recursively solve the subtasks. LLMs decompose with respect to the string concatenation operator. In other words, LLMs decompose the task of generating a sequence $t_1t_2\ldots t_n$ of
tokens to generating $t_1$ and then recursively generating $t_2\ldots t_n$ that are then combined using the string concatenation operator. This is not an ideal way to decompose the problem and then compose the solutions when the sequence of tokens represents a program. That is because a program has more than just a linear structure.
Symbolic solvers also decompose program synthesis task, but do so guided by the abstract syntax tree (AST) of the program. In other words, the pieces computed by the subproblems are not tokens, but ``subprograms'' that are composed using standard programming language operators, such as function composition. The approach we employ in this paper allows the LLM to compose it's outputs not by string concatenation, but by programming language operators. Since subprograms (as opposed to isolated tokens) have well-defined semantics, this also allows us to guide the LLM by giving it 
the semantics of the symbols it has generated. In totality, the ability to backtrack and repair (the prefix or suffix)
combined with the ability to compose subprograms together help make LLM generation less error prone, as we show in our evaluation.


\paragraph{Contributions.}
We make the following contributions in this paper.
We introduce a novel technique, \sysname, for solving programming by example (PBE) tasks using 
large language models (LLMs). \sysname\ blends ideas from symbolic forward and backward synthesis with
PBE synthesis using LLMs.
Our approach decomposes the program (being synthesized) with respect to operators that have a compositional semantics,
which allows us to compute and provide the semantics of one part of the program to the LLM when it is synthesizing the other part.
\sysname\ provides a mechanism to the LLMs to go back and repair components of the program it generated
 \emph{conditioned} on the semantics of the other components of the program.
We experimentally show that \sysname\ can solve task instances that are left unsolved even after multiple 
calls to an LLM that ask the LLM to fix the wrong program based on the error message~\cite{reflexion,llm-cegis}.
We also demonstrate the ability of \sysname\ to work across different target languages.
We provide insights behind the design of \sysname, which can inspire future explorations of LLM-supported
compositional program synthesis and approaches to exploit the semantics of program pieces in the process.
\endignore}


\def\invar{{\mathtt{x}}}
\def\outvar{{\mathtt{o}}}
\newcommand\MyCall[1]{{\textsc{#1}}}
\def\match{{\mathtt{match}}}
\def\re{{\mathtt{re}}}
\def\search{{\mathtt{search}}}
\def\group{{\mathtt{group}}}
\def\strip{{\mathtt{strip}}}

\newcommand\mycall[2]{{\textsc{#1}}}

\subsection{Motivating Example}

\begin{figure*}[t]
    \centering 
    \includegraphics[trim=0 50 0 0,clip,width=1.9\columnwidth]{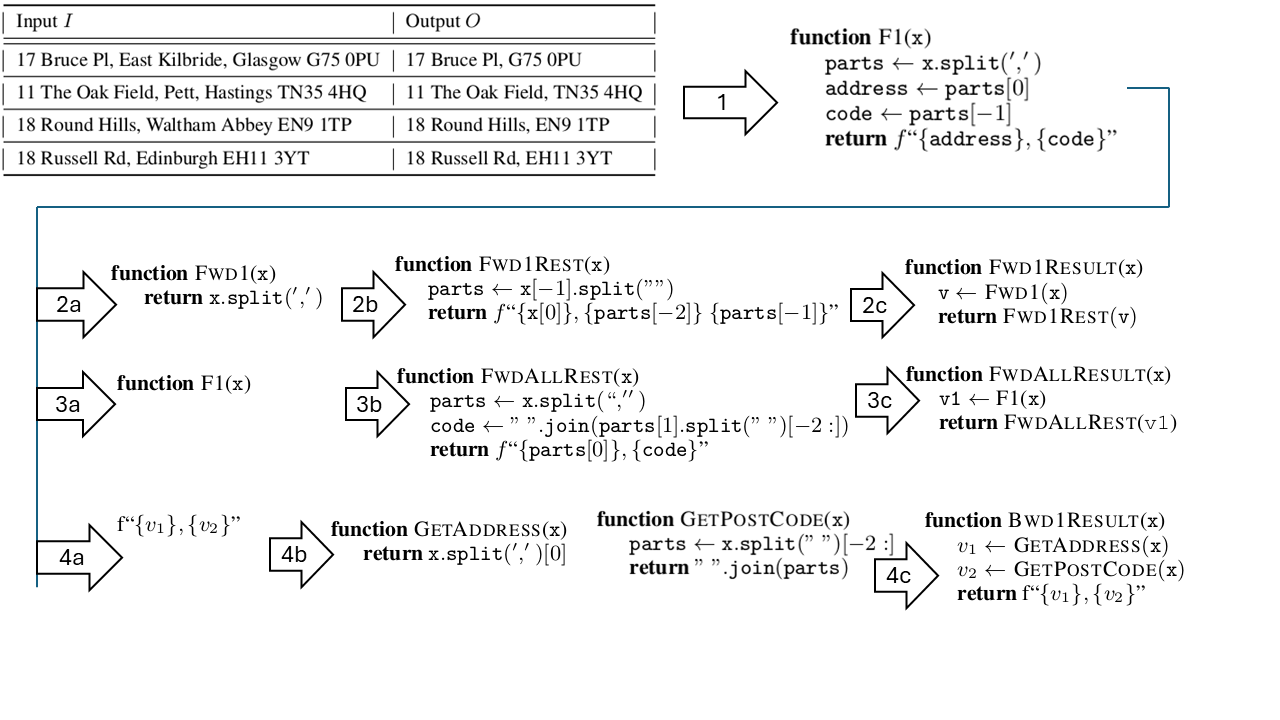}
    \caption{{\small{Illustration of our approach: We start with input-output (IO) examples. In Step (1), we use an LLM to generate a program $\mycall{F1}{}$. Since $\mycall{F1}{}$ does not generate the desired outputs, (2a) we decompose it and salvage $\mycall{Fwd1}{}$, (2b) we use LLM to solve the remaining task of transforming output of $\mycall{Fwd1}{}$ to $O$, and if successful, (2c) return the composed program. If unsuccessful, (3a) we salvage the full $\mycall{F1}{}$, (3b) use LLM to transform output of $\mycall{F1}{}$ to $O$, and if successful, (3c) return the composed program. If unsuccessful, (4a) we salvage the last step, (4b) use LLM to synthesize the two pieces needed to produce the 2 variables used in the last step, and (4c) return the composed program. There is a fourth if-then-else composition based strategy too in our approach.}}}\label{fig:motivating}
\end{figure*}

Consider the task of extracting the street name and postal code (in column $O$) from an address (in column $I$) as shown in  Figure~\ref{fig:motivating}.  The user has provided a few examples, consisting of inputs $I$ and corresponding outputs $O$. 
In Step~(1), we ask an LLM to synthesize a program for transforming $I$ to $O$.
The LLM returns the function $\Call{F1}{}$ shown in Figure~\ref{fig:motivating}.
This program is incorrect since it includes the city name in the output too.
We try a few decomposition strategies.

In strategy {\em{Forward1}},
we follow Steps (2a), (2b) and (2c).
In Step~(2a), the first operation in the incorrect program $\Call{F1}{}$, which splits the input string by a comma and space,
``, '', is salvaged as $\Call{Fwd1}{}$.
In Step~(2b) we execute $\Call{Fwd1}{}$ on the inputs $I$ to get some intermediate values $V$,
and ask the LLM to generate a program that converts $V$ to the final output $O$. 
The LLM returns $\Call{Fwd1Rest}{}$.
Finally, in Step~(2c), we compose the pieces to get the final program $\Call{Fwd1Result}{}$.

In strategy {\em{ForwardAll}},
we follow Steps (3a), (3b) and (3c).
In Step~(3a), the whole incorrect program $\Call{F1}{}$ is salvaged.
In Step~(3b), we execute $\Call{F1}{}$ on the inputs $I$ to get some intermediate values $V$,
and ask the LLM to generate a program that converts $V$ to the final output $O$. 
The LLM returns $\Call{FwdAllRest}{}$.
Finally, in Step~(3c), we compose the pieces to get the final program $\Call{FwdAllResult}{}$.

In strategy {\em{Backward1}},
we follow Steps (4a), (4b) and (4c).
In Step~(4a), the last line of the incorrect program that produces the output by joining two inputs with comma is salvaged.
In Step~(4b), we use the LLM to {\em{inverse execute}} the last line on the outputs $O$ to get some 
plausible choices for $v_1$ and $v_2$, 
and ask the LLM to generate two programs that convert $I$ to the found values for $v_1$ and $v_2$ respectively.
The LLM returns $\Call{GetAddress}{}$ and $\Call{GetPostCode}{}$.
Finally, in Step~(4c), we compose the pieces to get the final program $\Call{Bwd1Result}{}$.

A fourth strategy, {\em{IfThenElse}}, is used if the original program $\Call{F1}{}$ works correctly on some inputs but not all. This is not the case for our running example here. The first three strategies all worked on the running example, but that may not be the case always. We really need only any one strategy to succeed to declare success on the original task.

\ignore{
\begin{figure*}[t]
\small
\begin{tabular}{|l|l|}
\toprule
Input $I$ & Output $O$
\\ 
\midrule
\midrule
17 Bruce Pl, East Kilbride, Glasgow G75 0PU
&
17 Bruce Pl, G75 0PU
\\ \midrule
11 The Oak Field, Pett, Hastings TN35 4HQ
&
11 The Oak Field, TN35 4HQ
\\ \midrule 
18 Round Hills, Waltham Abbey EN9 1TP
&
18 Round Hills, EN9 1TP
\\ \midrule
18 Russell Rd, Edinburgh EH11 3YT
&
18 Russell Rd, EH11 3YT
\\
\bottomrule
\end{tabular}
\caption{Motivating task where given a column $I$ of addresses, a user wants to extract a new column of partial addresses shown in $O$.}\label{fig:motivating}
\end{figure*}

The user asks an LLM to synthesize a program for their task specified by the input-output
$(I, O)$ examples.
The LLM returns the function $\Call{F1}{}$ shown in Figure~\ref{fig:motivating-wrong-program}.
This program is incorrect since it includes the city name in the output too -- the incorrect
output generated by this program 
is shown in the table in Figure~\ref{fig:motivating-wrong-program} and the incorrect part
is underlined there.

\def\arga{\invar}
\def\parts{\mathtt{parts}}
\def\address{\mathtt{address}}
\def\code{\mathtt{code}}
\def\splita{\mathtt{split}}
\def\joina{\mathtt{join}}

\begin{figure*}[t]
\begin{tabular}{cc}
\begin{minipage}{0.45\textwidth}
    \small
    \begin{algorithmic}[H]
        \Function{F1}{$\arga$}
          \State $\parts \gets \arga.\splita(',')$
          \State $\address \gets \parts[0]$
          \State $\code \gets \parts[-1]$
          \State \Return $f$``$\{\address\}, \{\code\}$''
        \EndFunction
    \end{algorithmic}
\end{minipage}
    &
\begin{minipage}{0.45\textwidth}
    \small
\begin{tabular}{|l|}
\toprule
    Program Output $O' := \mathtt{F1}(I)$:
\\ 
\midrule
\small
17 Bruce Pl, {\underline{Glasgow }}G75 0PU
\\
    11 The Oak Field, {\underline{Hastings }}TN35 4HQ
\\
    18 Round Hills, {\underline{Waltham Abbey }}EN9 1TP
\\
    18 Russell Rd, {\underline{Edinburgh }}EH11 3YT
\\
\bottomrule
\end{tabular}
\end{minipage}
\end{tabular}
    \caption{Incorrect program \MyCall{F1} returned by the LLM and the output $O'$ it generates. 
    The incorrectly included city name in $O'$ are underlined.}
\label{fig:motivating-wrong-program}
\end{figure*}

How can we synthesize the correct program? The LLM-generated incorrect program 
in Figure~\ref{fig:motivating-wrong-program} has parts that are potentially reusable.
In fact, often either the {\em{whole program}} itself is almost correct and reusable,
or there is a {\em{prefix}} or {\em{suffix}} of the incorrect program that is reusable.
We present strategies that can be used to salvage parts of the incorrect program
and use them to build the correct program. 

\subsection{Forward1 Strategy}
We have observed that LLMs are often
able to find the correct initial operations to be performed on the input, {\em{even when they fail to
produce fully correct code}}. This is possibly due to the fact that the LLM has access to the inputs $I$ which helps in predicting the immediate operations to apply on $I$.
The forward1 strategy exploits the above observation to fix the wrong program as follows:
find the first operation(s) performed on the input by the incorrect program, execute the operations on each of the inputs $I$ in the IO examples to get new values $V$, create a new task of converting $V$ to $O$ and recursively solve the new problem.

In our running example, the first operation in the incorrect program $\Call{F1}{}$ splits the input string by a comma and space,
``, '', which is a reasonable first step. 
If we execute this first step on the inputs $I$, the intermediate values $V$ obtained are just lists of strings generated by
  applying $\mathtt{split}(``, '')$ to the inputs $I$. These values are shown in 
  Figure~\ref{fig:motivating-fwd1-program}. 
  We now ask the LLM to generate a program that converts $V$ to the final output $O$. The LLM returns the function 
  $\Call{Fwd1Rest}{}$, which transforms $V$ to $O$. The complete program $\Call{Fwd1Result}{}$ for IO examples $(I,O)$ can be obtained by composing the program $\Call{Fwd1}{}$ for the IO examples $(I,V)$  with the program 
  $\Call{Fwd1Rest}{}$ for the IO examples $(V,O)$ as shown in Figure~\ref{fig:motivating-fwd1-program}.

\begin{figure*}[t]
\begin{tabular}{cc}
\begin{minipage}{0.5\textwidth}
    \small
    \begin{algorithmic}[H]
        \Function{Fwd1}{$\arga$}
          \State \Return $\arga.\splita(',')$
        \EndFunction
        \Function{Fwd1Rest}{$\arga$}
          \State $\parts \gets \arga[-1].\splita(" ")$
          \State \Return $f$``$\{\arga[0]\}, \{\parts[-2]\}\ \{\parts[-1]\}$''
        \EndFunction
        \Function{Fwd1Result}{$\arga$}
          \State $\mathtt{v} \gets \Call{Fwd1}{\arga}$
          \State \Return $\Call{Fwd1Rest}{{\mathtt{v}}}$
        \EndFunction
    \end{algorithmic}
\end{minipage}
    &
\begin{minipage}{0.45\textwidth}
    \small
\begin{tabular}{|l|}
\toprule
    Output $V$ of $\Call{Fwd1}{I}$
\\ 
\midrule
\small
    {[17 Bruce Pl, East Kilbride, Glasgow G75 0PU]}
\\
    {[11 The Oak Field, Pett, Hastings TN35 4HQ]}
\\
    {[18 Round Hills, Waltham Abbey EN9 1TP]}
\\ 
    {[18 Russell Rd, Edinburgh EH11 3YT]}
\\
\bottomrule
\end{tabular}
\end{minipage}
\end{tabular}
    \caption{Program \MyCall{Fwd1Final} found by Forward1 Strategy and the intermediate values $V$ used to create the subproblem.}
\label{fig:motivating-fwd1-program}
\end{figure*}

\subsubsection*{Intuition behind Forward1} The key idea used in the Forward1 strategy is that of 
providing execution results to the LLM to help it complete a task. The subproblem in the Forward1
strategy is created by using the outputs of the executions of the initial operations as the new
inputs. Thus, the LLM now has knowledge of the correct new program state and it does not have to
complete the program just based on the original inputs and the partially-generated program.
In fact, we forget the original inputs and the partially-generated program, and let the LLM just start
fresh from the new program state.  This feedback to the LLM in the form of execution results helps
the LLM constantly correct its course and thus reduces the chances of mistakes made by the LLM.

\subsection{ForwardAll Strategy}

Sometimes, the LLM is able to generate
{\em{almost}} correct programs and only simple tweaks are needed to fix the output of the
almost-correct program. Some examples here are cases when the original LLM-generated program misses 
formatting the output correctly or generates an output that has some extraneous characters. In these cases, 
the ForwardAll strategy gives the LLM a second chance to postprocess the output of the original wrong
program, say by applying the required trimming or 
formatting operations. In our running example,
we observe that the original incorrect program has extra characters in the output and we could try
to see if a second LLM call can remove them.

On our running example, the ForwardAll strategy works by first executing the wrong program $\Call{F1}{}$ 
  (shown in Figure~\ref{fig:motivating-wrong-program})
on the original inputs $I$. The execution of $\Call{F1}{}$ is successful on all inputs $I$
and it generates the outputs $O' := \Call{F1}{I}$ as shown in Figure~\ref{fig:motivating-wrong-program}.
The ForwardAll strategy tries to generate a correct program by generating a program that
would transform the intermediate values $O'$ to the desired output $O$.
The program returned by the LLM for the subtask of transforming $O'$ to $O$ is shown 
in Figure~\ref{fig:motivating-fwdall-program} as function $\Call{FwdAllRest}{}$.
The final correct program $\Call{FwdAllResult}{}$ is obtained by simply composing the wrong program
$\Call{F1}{}$ with $\Call{FwdAllRest}{}$.

\begin{figure*}[t]
\begin{tabular}{c|c}
\begin{minipage}{0.55\textwidth}
    \small
    \begin{algorithmic}[H]
        \Function{FwdAllRest}{$\arga$}
          \State $\parts \gets \arga.\splita(``, '')$
          \State $\code \gets "\ ".\joina(\parts[1].\splita("\ ")[-2:])$
          \State \Return $f$``$\{\parts[0]\}, \{\code\}$''
        \EndFunction
    \end{algorithmic}
\end{minipage}
    &
\begin{minipage}{0.35\textwidth}
    \small
    \begin{algorithmic}[H]
        \Function{FwdAllResult}{$\arga$}
          \State $\mathtt{v1} \gets$ \Call{F1}{$\arga$}
          \State \Return \Call{FwdAllRest}{\texttt{v1}}
        \EndFunction
    \end{algorithmic}
\end{minipage}
\end{tabular}
    \caption{Program \MyCall{FwdAllFinal} found by the ForwardAll Strategy.}
\label{fig:motivating-fwdall-program}
\end{figure*}

\subsubsection*{Intuition behind ForwardAll}
Apart from the benefits from providing execution results mentioned above,
we also note that Forward1 and ForwardAll are creating simpler subproblems
for the LLM to solve. A fairly popular and common approach for improving robustness
of LLMs, especially for code generation, is that of repeatedly invoking the LLM 
asking it to fix its prediction by providing information related to the mistake
in its prediction. We will call this strategy ``counter-example guided inductive
synthesis'', or \emph{CEGIS.} The ForwardAll strategy may look like CEGIS at first
glance. However, there is an important difference: in all our strategies, including
ForwardAll, we are having the LLM solve new simpler subtasks, and we completely 
forget the original task. In contrast, CEGIS keeps the task in tact, and provides
more and more information iteratively in a failure-guided fashion.

\subsection{Backward1 Strategy}

The third strategy for synthesizing a correct program from the given input-output examples
is based on keeping the ``suffix'' of the incorrect program intact and adding a fixed ``prefix'' to it. 
The intuition behind the Backward1 strategy is that LLMs are also highly likely to correctly predict the last operations
required for converting inputs $I$ to outputs $O$. The reason is that the outputs $O$ are available to the LLM, and 
it is often possible to predict the operation required to create $O$ by looking at the structure of $O$.
The Backward1 strategy works by 
(1) extracting the last operation used in the wrong program, 
(2) predicting the inputs, say $I_1$ and $I_2$, required by the last (say binary) operation to generate the actual outputs $O$,
and
(3) creating two PBE subproblems - one to convert $I$ to $I_1$ and another to convert $I$ to $I_2$, and solving the
two subproblems.

\begin{figure*}[t]
\begin{tabular}{c|c}
\begin{minipage}{0.40\textwidth}
    \small
\begin{tabular}{|l|l|}
\toprule
    Predicted $\mathtt{address}$
    &
    Predicted $\mathtt{postcode}$
\\ 
\midrule
\small
    17 Bruce Pl & G75 0PU
\\
    11 The Oak Field & TN35 4HQ
\\
    18 Round Hills & EN9 1TP
\\ 
    18 Russell Rd & EH11 3YT
\\
\bottomrule
\end{tabular}
\end{minipage}
    &
\begin{minipage}{0.55\textwidth}
    \small
    \begin{algorithmic}[H]
        \Function{GetAddress}{$\arga$}
          \State \Return $\arga.\splita(',')[0]$
        \EndFunction
        \Function{GetPostCode}{$\arga$}
          \State $\parts \gets \arga.\splita("\ ")[-2:]$
          \State \Return $"\ ".\joina(\parts)$
        \EndFunction
        \Function{Bwd1Result}{$\arga$}
          \State $v_1 \gets \Call{GetAddress}{\arga}$
          \State $v_2 \gets \Call{GetPostCode}{\arga}$
          \State \Return f``$\{v_1\}, \{v_2\}$''
        \EndFunction
    \end{algorithmic}
\end{minipage}
\end{tabular}
\caption{The predicted inputs for the last step is shown on the left. The two programs that generate those two inputs from the original input are shown on the right, followed by the final correct program obtained by the Backward1 Strategy.}
\label{fig:motivating-bwd1-program}
\end{figure*}

On our running example, the
last operation that is used to compute the return value of the (incorrect) function
$\Call{F1}{}$ is a string concatenation operation,
{\texttt{f"\{address\},\ \{postcode\}"}}, which concatenates two strings
stored in the variables {\texttt{address}} and {\texttt{postcode}} with a 
separator containing a comma and a space.
Next, we need to apply the ``inverse semantics'' of this binary concatenation operation to find
the two inputs that would generate the outputs $O$.
We can use the LLM to predict these values, which we can then verify using the usual operational
semantics of the concatenation operator. The LLM is able to successfully predict these values for
our running example, which are shown in
Figure~\ref{fig:motivating-bwd1-program} on the left.
This gives us two subproblems: converting the original inputs $I$ to the street address, and converting
the original inputs $I$ to the postcode.
These two subproblems are simpler than the original problem. We ask the LLM to solve the two subproblems.
If the LLM is able to successfully solve the two problems, say by returning the functions
$\Call{GetAddress}{}$ and $\Call{GetPostCode}{}$ shown in Figure~\ref{fig:motivating-bwd1-program},
then we can return the final correct program by executing these two programs (in parallel) and
generating the final output by applying the concatenation operator from the wrong program.

\ignore{
\noindent\begin{minipage}{\textwidth}
  \centering
  \begin{minipage}{.45\textwidth}
    \centering
    \captionof{algorithm}{test algorithm 1}
    \label{alg:alg1}
    \begin{algorithmic}
    \WHILE{$N \neq 0$}
    \IF{$N$ is even}
    \STATE $X \Leftarrow X \times X$
    \STATE $N \Leftarrow N / 2$
    \ENDIF
    \ENDWHILE
    \end{algorithmic}
  \end{minipage}
  \begin{minipage}{.45\textwidth}
    \centering
    \captionof{algorithm}{test algorithm 2}
    \label{alg:alg2}
    \begin{algorithmic}
    \WHILE{$N \neq 0$}
    \IF{$N$ is even}
    \STATE $X \Leftarrow X \times X$
    \STATE $N \Leftarrow N / 2$
    \ENDIF
    \ENDWHILE
    \end{algorithmic}
  \end{minipage}
  \captionof{figure}{Two algorithms side by side}
  \label{fig:twoalg}
\end{minipage}
\endignore}

\subsubsection*{Intuition behind Backward1}
Backward1 also shares benefits of exploiting the execution semantics and performing
problem decomposition into simpler subproblems mentioned above. Specifically, we note
that Backward1 can create multiple small subproblems. Additionally, the Backward1
strategy also inherits the benefits of (symbolic) backward synthesis~\cite{gulwani2011automating}.
In particular, in contrast to forward synthesis, backward synthesis is goal-directed
and looking at the goal (output) helps with limiting the search space in symbolic 
backward synthesis, which also potentially plays a role in our Backward1 strategy.

\subsection{IfThenElse Strategy}

A final strategy for generating a correct program starting with an incorrect one is the
IfThenElse strategy where we use the incorrect program in one branch of an if-then-else
program and synthesize the condition and the program for the other branch.
This can be used when the LLM generates a program
that works on {\em{some}} inputs correctly, but does not work on some others. 

The IfThenElse strategy works as follows. First, the LLM-generated program $\Call{F1}{}$ is executed on all
the inputs $I$ to generate outputs $O'$. Next we partition the input set $I$ into $I_1$ and $I_2$ where
$I_1 = \{i \in I\mid \Call{F1}{i} = o\}$ contains all the inputs where $\Call{F1}{}$ produces the correct outputs.
The subtask now is to synthesize a program that works correctly on $I_2$. If we are able to find such a program
$\Call{F2}{}$, then we generate a condition $c$ that is true for $I_1$ but false for $I_2$. 
If we find such a condition $c$, the overall correct program is obtained
as ``{\texttt{if (c(i)) then \Call{F1}{i} else \Call{F2}{i}}}''. 
In practice, we also use the intermediate values generated in $\Call{F1}{i}$, apart from $i$ itself, 
to find the required condition.
We will discuss more on condition generation later.

\begin{table*}[t]
\small
\begin{tabular}{|l|l|l|}
\toprule
    Input $I$ & Output $O$ & $I.\splita(",\ ")[-2]$
\\ 
\midrule
17 Bruce Pl, East Kilbride, Glasgow G75 0PU
&
East Kilbride
&
East Kilbride
\\ 
11 The Oak Field, Pett, Hastings TN35 4HQ
&
Pett
&
Pett
\\ 
18 Round Hills, Waltham Abbey EN9 1TP
&
Waltham Abbey
&
18 Round Hills
\\ 
18 Russell Rd, Edinburgh EH11 3YT
&
Edinburgh
&
18 Russell Rd
\\ 
49 Kent Dr, Redwood City, CA 94025
&
Redwood City
&
Redwood City
\\ 
1229 237th Pl NE, Kirkland, WA 98075
& Kirkland
& Kirkland
\\
\bottomrule
\end{tabular}
\caption{The input-output examples for a task of extracting city names from addresses, and the 
    output of the incorrect program generated by the LLM. The incorrect program is shown in the 
    header of third column.}\label{tab:ite}
\end{table*}

For our running example, the program $\Call{F1}{}$ works incorrectly on all inputs in $I$, and hence the 
IfThenElse strategy is not applicable. Instead consider a slightly modified version of the task shown
in the first two columns of Table~\ref{tab:ite} where we are extracting city names from the addresses. 
We added two more addresses
in the input set $I$ that are US addresses and have a different format compared to the first four.
The LLM generates the program $\Call{F1}{}$ (shown in Figure~\ref{fig:ite-all}) that returns 
$\arga.\splita(",\ ")[-2]$, and the output of this program is 
shown in the third column in Table~\ref{tab:ite}.
Clearly, the LLM focuses more on the first and last addresses and returns a program that works for the first
two and the last two inputs, but does not work for the middle two inputs.
This is again a common occurrence: if the uncommon case is not at the end of the list of IO examples, then 
LLMs give less attention to them and end up generating programs that do not work for those cases.

\begin{figure*}[t]
\begin{tabular}{c|c}
\begin{minipage}{0.65\textwidth}
    \small
    \begin{algorithmic}[H]
        \Function{F1}{$\arga$}
          \State $\parts \gets \arga.\splita(',\ ')$
          \State \Return $\parts[-2]$
        \EndFunction
        \Function{F2}{$\arga$}
          \State $r \gets \verb|r',\s*([^,]+)\s+[A-Z]{1,2}\d{1,2}\s+\d[A-Z]{2}'|$
          \State $\match \gets \re.\search(r, \arga)$
          \State \Return $\match.\group(1).\strip()$
        \EndFunction
    \end{algorithmic}
\end{minipage}
    &
\begin{minipage}{0.35\textwidth}
    \small
    \begin{algorithmic}[H]
        \Function{F3}{$\arga$}
          \State $\parts \gets \arga.\splita(',\ ')$
          \State $v_1 \gets \parts[-2]$
          \If{$\mathtt{len}(\parts)$==$2$}
            \State \Return $v_1$
          \EndIf
          \State \Return $\Call{F2}{\arga}$
        \EndFunction
    \end{algorithmic}
\end{minipage}
\end{tabular}
    \caption{The programs $\MyCall{F1}$ and $\MyCall{F2}$ that work on different inputs, but together cover all inputs, and the final program $\MyCall{F3}$ obtained by using one or the other depending on a condition (on an intermediate value created in $\MyCall{F1}$.}
\label{fig:ite-all}
\end{figure*}

Since the program $\Call{F1}{}$ does not work on the 3rd and 4th inputs, we make a 
second LLM call to find a program that works on only those two inputs. The LLM returns
the program $\Call{F2}{}$ shown in Figure~\ref{fig:ite-all}. The final task is to generate
a condition to use to decide which program to use on any given input. We use the LLM to generate
this condition as well. Rather than asking the LLM to find a condition on the inputs, we
present to the LLM a program state reached by executing $\Call{F1}$ on the input; thus, the LLM
can use intermediate variables, such as $\parts$, in the condition it generates.
The final program obtained by putting together the condition and the programs $\Call{F1}{}$ and $\Call{F2}{}$
is shown as program $\Call{F3}{}$ in Figure~\ref{fig:ite-all}.
Note that we assumed that executing $\Call{F1}{}$ on all inputs leads to a program state. If it is the 
case that some inputs throw an exception, then we can only use the input values to learn a condition
unless $\Call{F1}{}$ throws an exception on {\em{all}} inputs in the other class, in which case we can
combine $\Call{F1}{}$ and $\Call{F2}{}$ using a $\mathtt{try-catch}$ block.

\subsubsection*{Intuition behind IfThenElse}
The motivation for the IfThenElse strategy comes from the observation that LLMs have a tendency
to ignore certain input-output examples when they are presented a set of IO examples. Typically,
the IO examples in the middle of a list tend to have less influence on the LLM generation compared to 
the IO examples at the start or end of the list. This causes the LLM to generate programs that work
on some, but not all, inputs. Our strategies also give the LLM a chance to ``backtrack'' and revisit
some parts of the programs -- and add an ``conditional branch'' branch, for example. 
This is not possible to do in the traditional linear (left to right, top to bottom) generation of
code performed by the LLMs.

\endignore}

\def\st{{\mathtt{st}}}
\def\retvar{{\mathtt{ret}}}
\def\val{{\mathtt{v}}}
\def\inval{{\mathtt{v}_{in}}}
\def\outval{{\mathtt{v}_{out}}}
\def\Values{{\mathtt{Values}}}
\def\Example{{\mathtt{ex}}}
\def\Examples{{\mathtt{Ex}}}
\def\PL{{\mathtt{PL}}}
\def\Programs{{\mathtt{Prgms}}}
\def\program{{\mathtt{F}}}
\newcommand\Sem[1]{{[\![{#1}]\!]}}
\newcommand\synconfig[1]{{\ensuremath{\langle{{#1}}\rangle}}}

\section{Programming by Example}

We now define the problem of programming-by-example.
%

Let $\Values$ be the universal set of (concrete) values, which includes strings, numbers, lists, and tuples. A value in $\Values$ will be denoted by $\val$, but it may be annotated to
indicate if it is an input value, $\inval$, or an output value, $\outval$. 
Let $\Programs$ be the set of programs in a programming language. The semantics of a program is a function that maps (input) values to (output) values. We do not differentiate between the syntax $\program$ of a program and its semantics, so $\program$ will denote a program, and $\program(\val)$ will denote the output of the program $\program$ when it is applied to the input value $\val\in\Values$. Since values can be tuples, our programs can can take multiple inputs (as a tuple) and produce multiple outputs (as a tuple). Since we are not differentiating between syntax and semantics, we will use $\val$ also as a program variable.

An {\em{IO example}}, or simply an {\em{example}}, $\Example$ is a pair $(\inval, \outval)$ consisting 
of an input value $\inval$ and an output value $\outval$.
A \emph{PBE task} is a set $\Examples$ of examples.
A {\em{solution}} of the synthesis task $\Examples$ in the target language $\Programs$
is a program $\program\in\Programs$ such that $\program(\inval) = \outval$ for every
example $(\inval,\outval)\in\Examples$.
If program $\program$ \emph{solves} the PBE task $\Examples$, we denote it as
$\program \models \Examples$.
The notation $\program \not\models \Examples$ denotes the negation of $\program\models\Examples$.

%
%
%
%

\begin{example}\label{ex:synthesistask}
    Consider the motivating scenario shown in Figure~\ref{fig:motivating}.
    Here the PBE task $\Examples$ contains 4 examples. The input values are strings and the output values are strings. 
    In the first example $(\inval_1, \outval_1)$, the input $\inval_1$ is 
    ``17 Bruce Pl, East Kilbride, Glasgow G75 0PU''
    and the output value $\outval_1$ is the string ``17 Bruce Pl, G75 0PU''.
    The Python program $\Call{F1}{}$ in Figure~\ref{fig:motivating} is not 
    a solution for this PBE task, but the Python program $\Call{Fwd1Result}{}$ in 
    the same figure is a (correct) solution.
  \qed
\end{example}

We assume that the set $\Programs$ of programs in the language $\PL$
is closed under the usual composition operators, such as 
sequential composition, which we used to combine small programs into a
large program in Steps~(2c), (3c) and (4c) in Figure~\ref{fig:motivating}),
and the if-then-else operator.

\ignore{  
\subsection{Compositions of Programs}
\label{sec:composition}

\begin{figure}[t]
\begin{center}
\centerline{\includegraphics[width=0.9\columnwidth,trim=0 335 100 50,clip]{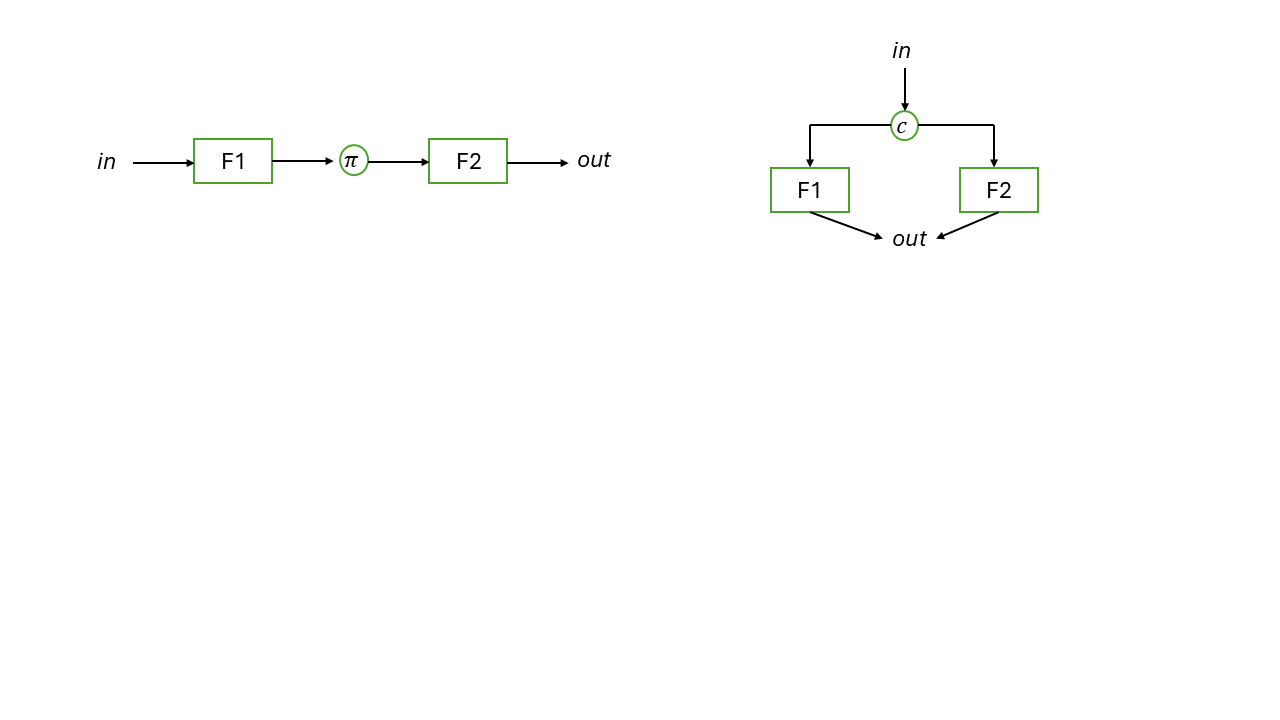}}
\caption{\footnotesize{Two composition operators on programs whose semantics can be defined by composing semantics of the parts. Functions are shown to take one input for simplicity; they can take multiple inputs in general.}}
\label{fig:compositions}
\end{center}
\end{figure}

Programs are structured sequences of text whose semantics is defined by defining semantics of the parts
and composing them together. There are two particular composition operators on programs of interest for
this work: a sequential composition operator and a parallel composition operator; see Figure~\ref{fig:compositions}.
We assume that the set $\Programs$ is closed under these two composition operators.

The \emph{sequential composition operator}, $\circ$, is a binary operator that maps a tuple
$(\Programs_1,\program_2)$, where $\Programs_1$ is an 
ordered sequence of programs, $\langle \program_{11},\ldots,\program_{1n}\rangle$,
to a program, written as $\Programs_1 \circ \program_2$.
The semantics of $\circ$ is given as follows: For any program state $\st$,
\begin{eqnarray}
    \Sem{\Programs_1 \circ \program_2}(\st) & := & \Sem{\program_2}(
         \biguplus_i  \Sem{\program_{1i}}(\st) )
\end{eqnarray}
where $\uplus_i \Sem{\program_{1i}}(\st)$ denotes the state
       $\langle\invar_1 \mapsto \Sem{\program_{11}}(\st)(\retvar), \ldots,
                \invar_n \mapsto \Sem{\program_{1n}}(\st)(\retvar) \rangle$.

We assume that our target language (for synthesis),
$\PL = (\Programs, \Sem{\cdot})$, is closed under $\circ$; that is,
for any sequence $\Programs_1$ and a program $\program_2$,
the program $\Programs_1 \circ \program_2$ is also a program in $\Programs$.
For simplicity, and without any loss of generality, whenever we are discussing
sequential composition, we will assume that
$\Programs_1$ has exactly two programs $\langle \program_{11},\program_{12}\rangle$,
and the final composed $\program$ has two inputs $\invar_1$ and $\invar_2$.
In this case, the sequential composition is represented as
$\lambda{\invar_1,\invar_2}: \program_2( \program_{11}(\invar_1, \invar_2), \program_{12}(\invar_1, \invar_2) )$.


The parallel composition operator, $\parallel_c$, is a binary operator on programs  parameterized
by a  condition $c$ which is a mapping from program states to Boolean values. We assume the following:
\begin{itemize}
    \item  Closure of $\Programs$ under $\parallel_c$: For any two programs $F_1, F_2\in\Programs$ and 
        a mapping $c$, the program $F_1 \parallel_c F_2$ is also a program in $\Programs$.
    \item Semantics of $\parallel_c$: For any program state $\st$,
\begin{eqnarray}
    \Sem{F_1 \parallel_c F_2}(\st) & := & \left\{
        \begin{array}{ll}
            \Sem{F_1}(\st) & \mbox{if $c(\Sem{F_1}(\st))$ is true}
            \\
            \Sem{F_2}(\st) & \mbox{otherwise}
        \end{array}\right.
        \label{eqn:ite-semantics}
\end{eqnarray}
\end{itemize}
Note that the condition $c$ is evaluated on the program state produced by $\Sem{F_1}(\st)$. 
Alternatively, we can also consider the case when the condition $c$ is applied directly on $\st$ (the input state).
The version used in Equation~\ref{eqn:ite-semantics} allows the condition $c$ to be based on intermediate values computed by
$F_1$ (apart from the inputs). We experiment with both notions, but just use the single notation $\parallel_c$ to cover
both versions. There is another parallel composition, $\parallel$, used in experiments where we use a try-catch
block to combine the two parts. In our formalism, it's semantics can be defined as:
\begin{eqnarray}
    \Sem{F_1 \parallel F_2}(\st) & := & \left\{
        \begin{array}{ll}
            \Sem{F_1}(\st) & \mbox{if $\Sem{F_1}(\st)\neq\bot$}
            \\
            \Sem{F_2}(\st) & \mbox{otherwise}
        \end{array}\right.
        \label{eqn:try-catch-semantics}
\end{eqnarray}
Recall that $\bot$ denotes that an undefined value and $\Sem{F_1}(\st)$ will equal $\bot$ when,
for example, the program $F_1$ throws an exception when executed on $\st$.
For simplicity, we will limit the discussion to the two composition operators, $\circ$ and $\parallel_c$, but 
all the development carry over to the case when we also have other variants of the sequential and parallel composition operators.

We note that the semantics of the composition of programs is defined as a composition of the semantics of the
parts. Thus, one could use the semantics of one part to influence the generation of the other part if the two
parts are combined by one of these two composition operators.

\begin{example}[Compositions]
    Consider the program $\Call{FwdAllResult}{}$ shown in Figure~\ref{fig:motivating-fwdall-program}.
    It is a sequential composition, $\langle \Call{F1}{} \rangle \circ \Call{FwdAllRest}{}$,
    of the program $\Call{F1}{}$ in Figure~\ref{fig:motivating-wrong-program} and
    the program $\Call{FwdAllRest}{}$ in Figure~\ref{fig:motivating-fwdall-program}.

    Consider the program $\Call{F3}{}$ shown in Figure~\ref{fig:ite-all}. 
    It is obtained by parallel composition,
    $\Call{F1}{} \parallel_c \Call{F2}{}$, of $\Call{F1}{}$ and $\Call{F2}{}$ in the same figure, where
    the condition $c$ checks that $\mathtt{len}(\mathtt{parts}) == 2$. Note that the value of the variable
    $\mathtt{parts}$ is computed by $\Call{F1}{}$ and is not part of the input. \qed
\end{example}

\subsection{Large Language Models}

Large Language Models (LLMs) are at the core of our approach for solving PBE tasks.
LLMs are huge neural networks trained on the next-token prediction task.
A token is a part of a word that occurs commonly in text. The training data for LLMs contains
text sourced from the internet that has been broken down into tokens. If $t_0,t_1,t_2,\ldots$
is the tokenization of a piece of text in the training dataset, then an LLM is trained to predict
every $t_k$ conditioned on the previous $N$ tokens, $t_{k-N},t_{k-N+1},\ldots,t_{k-1}$. In other 
words, an LLM is learning the following probability:
$$\mathtt{Prob}(t_k \mid t_{k-N},t_{k-N+1},\ldots,t_{k-1})$$

The LLM is given a prompt, which is just a starting sequence of tokens, and the model is used iteratively 
to predict the next token, conditioned on the prompt and the previously predicted tokens. 
Thus, an LLM can {\em{complete}} a sequence of words by returning another sequence of words. 
Note that the tokens returned by the LLM are concatenated with the previously returned tokens.
In other words, the LLM {\em{decomposes}} the task of predicting a sequence $t_k,t_{k+1},\ldots$ of tokens to predicting the
first token $t_k$ and recursively solving the subtask of predicting the sequence $t_{k+1},t_{k+2},\ldots$.
The final prediction is obtained by {\em{concatenating}} the predicted first token with the solution to the subtask -- thus, the (de)composition operator used by LLMs is string concatenation.

Symbolic synthesis approaches also \emph{decompose} the task of generating a program to the task of predicting 
subprograms. 
Forward synthesis, which is popularly known as enumerative or bottom-up synthesis, works by first generating
the ``prefix'' of the program -- part of the program that works directly on the inputs -- fixing that prefix
and then generating the remaining program recursively. The two parts are composed using sequential composition 
to construct the final program.
Backward synthesis, which is popularly known as top-down synthesis, works by first generating
the ``suffix'' of the program -- part of the program that directly produces the output -- fixing that suffix
and then generating the remaining program recursively. The two parts are again composed using sequential composition 
to construct the final program.
In general, symbolic synthesizer generate subprograms that are 
composed using standard program composition operators, such as the sequential and parallel composition operators described
above.

When used for program generation or synthesis tasks, LLMs view programs as strings and use the same strategy of
decomposing the synthesis problem into the subtasks of generating the token. This works reasonably well for 
generating small programs in popular programming languages, but the probability of getting correct completions
drops as the program length increases, and it also drops as we move to less popular target languages.
Individual tokens are unlikely to have any semantics. 
Unlike LLMs, symbolic synthesizer decompose the PBE task into subtasks at well-defined program boundaries, yielding
parts that have well-defined semantics. 
In this paper, we present an approach that combines LLM generation of tokens with symbolic synthesizers way of 
composing programs.
Our main hypothesis is that
{\em{if we decompose programs in a way that the program pieces
have well-defined semantics, then we can present the semantics of the partially generated programs to the LLM,
which can help improve the probability of LLM generating correct code overall.}}

\endignore}

\section{\sysname: Failure-Guided Compositional Program Synthesis}

In this section, we present our approach for synthesizing a program that is consistent with
a given set of input-output examples.

%

\ignore{
The basis for our approach can be described using a small set of simple inference rules. 
The rules are presented in Figure~\ref{fig:forward-backward}.
Each rule presents one \emph{sound} way to establish $F\models \Examples$. 
We will later discuss how we turn these rules into a synthesis approach.

\begin{figure}[t]
  \[
      \begin{array}{l@{\quad}l}
  \inferrule[Check]
          { }
    {F \models \Examples}
      &
          {\mbox{if } \Sem{F}(\st)(\retvar) = v \mbox{ for all }(\st,\val)\in\Examples}
      \\*[2em]
  \inferrule[Forward]
    { 
      F_2 \models \Examples'}
          {\langle F_{11},F_{12}\rangle \circ F_2 \models \Examples}
      &
          {\mbox{if } \Examples' = \{ (\uplus_i \Sem{F_{1i}}(\st),  v) \mid (\st,v)\in\Examples\} }
      \\*[2em]
    \inferrule[Backward]
    { 
          F_{11} \models \Examples_1 \\ 
          F_{12} \models \Examples_2 \\ 
          }
          {\langle F_{11},F_{12}\rangle \circ F_2 \models \Examples}
      &
          {\mbox{if } \Examples_i = \{ (\st,\st'|_i) \mid (\st,v)\in\Examples, v = \Sem{F_2}(\st')(\retvar)\} }
      \\*[2em]
          \inferrule[IfThenElse]
    { F_1 \models \Examples_1 \\
          F_2 \models \Examples_2 }
          {F_1 \parallel_c F_2 \models \Examples_1\cup\Examples_2}
      &
          {\mbox{if }\forall{(\st,\val)\in\Examples_1}: c(\Sem{F_1}(\st))\;\wedge\; 
           \forall{(\st,\val)\in\Examples_2}: \neg c(\Sem{F_1}(\st))}
  \end{array}
\]
    \caption{Inference rules for forward, backward, and if-then-else program synthesis.}
  \label{fig:forward-backward}
\end{figure}

\emph{Inference rule~\textsc{Check}}: This rule is an axiom for establishing $F\models\Examples$ using it's definition,
which is stated in the side condition.
The side condition $\forall{(\st,\val)\in\Examples}: \Sem{F}(\st)(\retvar) = \val$ for applicability 
of the inference rule can be discharged
by just executing $F$ on the initial state $\st$ and checking if the value of $\retvar$ in the
end program state is equal to $\val$, for every $(\st,\val)\in\Examples$.

\emph{Inference rule~\textsc{Forward}}: This rule establishes $\langle F_{11},F_{12}\rangle \circ F_2\models \Examples$
by reducing it to a PBE problem $\Examples'$ that is solved by $F_2$.
More specifically,
the inference rule~\textsc{Forward} states that
\emph{if} 
$F_2$ solves the PBE (sub)task $\Examples'$, and 
the new set of examples $\Examples'$ is created from $\Examples$ by
including $(\st',\val)$ in $\Examples'$ for every $(\st,\val)\in\Examples$, 
where $\st'$ is a new state containing the outputs (return values) of both
$F_{11}$ and $F_{12}$ on $\st$, 
\emph{then} we can assert that $\langle F_{11},F_{12}\rangle \circ F_2 \models \Examples$.
We make the following observations about the \textsc{Forward} inference rule:
\\
The \textsc{Forward} rule proves $F\models \Examples$ by decomposing $F$ into
$\langle F_{11},F_{12}\rangle \circ F_2$.
\\
(2) The program $F$ can be decomposed into $\Programs_1\circ F_2$ in many different ways
and the rule applies to \emph{any} such decomposition and the rule succeeds when 
$F_2$ solves a subtask $\Examples'$ that is generated using $F_{11},F_{12}$ and $\Examples$.
\\
(3) The premise $F_2\models \Examples'$ can be recursively established by any of the inference rules.
\\
(4) The new PBE task $\Examples'$ is generated by forward executing $F_{11}$ and $F_{12}$ on the 
inputs, and hence the name \textsc{Forward}.

\emph{Inference rule \textsc{Backward}}: This rule is similar to the \textsc{Forward} rule except that it reduces
the proof of $\langle F_{11},F_{12}\rangle \circ F_2 \models \Examples$ to the proof that 
$F_{11}$ and $F_{12}$ solve the appropriately generated 
PBE subtasks $\Examples_1$ and $\Examples_2$ respectively.
The observations made above for the ~\textsc{Forward} rule have their corresponding analogues
for the \textsc{Backward} rule too.
In particular, note that the new PBE subtasks $\Examples_1$ and $\Examples_2$ can only be obtained by ``inverse execution''
of $F_2$; that is, by finding the state $\st'$ that will produce the desired output $\val$ using
the semantics of $F_2$, and hence the name \textsc{Backward}.
Thus, both the \textsc{Forward} and \textsc{Backward} rules decompose the proof of $F\models\Examples$
to proofs of subclaims $F' \models \Examples'$ for some $F'$ and $\Examples'$.

\emph{Inference rule \textsc{IfThenElse}}: This inference rule establishes when an if-then-else
composition, $F_1 \parallel_{c} F_2$, solves a PBE task $\Examples$. As expected, it states that
the set $\Examples$ should partition into $\Examples_1\cup\Examples_2$ such that $F_1$ solves
$\Examples_1$, $F_2$ solves $\Examples_2$, and the condition $c$ should distinguish the two sets
of end states obtained by applying $F_1$ on the input states of the two example sets.

\begin{proposition}[Soundness]\label{prop:soundness}
If there is a derivation of $F\models\Examples$ using the inference rules then
it is indeed the case that $F\models\Examples$.
\end{proposition}

The inference rules in Figure~\ref{fig:forward-backward} are not interesting for use in
verification since given $F$ and $\Examples$, it is easy to verify if $F\models\Examples$
by just executing $F$ on each input state in $\Examples$.
However, we can use them to guide \emph{synthesis} using an LLM as an oracle to supply the program
$F$ used in the rules, which we discuss next.

\subsection{Failure-Guided Synthesis}

We now develop our failure-guided synthesis procedure by exploiting the inference rules.
Starting with only the set $\Examples$ of input-output examples, the main idea is to synthesize the program $F$
that solves this PBE task by attempting to find a proof for $F\models\Examples$ for some $F$ using the inference
rules.

\endignore}

Consider the PBE task given as the set $\Examples$ of input-output examples.
The \sysname\ approach solves the PBE task as follows:
\begin{itemize}
    \item Use an LLM with an appropriate prompt to generate a candidate program $\program$ in $\Programs$
    \item Execute $\program$ on the input values in $\Examples$ to check if 
        $\program\models\Examples$
    \item if $\program\models\Examples$, return $\program$
    \item Compute the subset $\Examples_1$ of examples on which $\program$ works correctly; that is,
        $$
           \Examples_1 = \{ (\inval_i,\outval_i)\in\Examples \mid \program(\inval_i) = \outval_i \}
        $$
        Given $\Examples$, $\program$, and $\Examples_1$, we then use one of the following four strategies to complete the task.
\end{itemize}
        
\paragraph{IfThenElse strategy:} If $\Examples_1$ is not empty, then we compute $\Examples_2 = \Examples \setminus \Examples_1$ and 
                we recursively solve
                the PBE task $\Examples_2$ to find $\program_2$ 
                such that $\program_2\models\Examples_2$.
                If we are successful in finding $\program_2$, we use an LLM to synthesize the condition $c$ such that 
                $c(\inval)$ is true for all $(\inval,\_)\in\Examples_1$
                and
                $c(\inval)$ is false for all $(\inval,\_)\in\Examples_2$.
                If we are successful in finding such a condition $c$, we return the program:
    \begin{algorithmic}[H]
        \Function{ITE}{$\val$}
          \If{ $c(\val)$ is true}
           \State \Return $\program(\val)$
          \Else
           \State \Return $\program_2(\val)$
          \EndIf
        \EndFunction
    \end{algorithmic}
                We will discuss condition synthesis, and its possible extensions, in more detail later.

            \paragraph{ForwardAll strategy:} We construct new IO examples using the value computed by $\program$ as the input and the original outputs as the desired outputs; that is,
        $$
                \Examples' = \{ (\val'_i,\outval_i) \mid (\inval_i,\outval_i)\in\Examples, \val'_i=\program(\inval_i)\}
        $$
         We recursively solve the PBE task $\Examples'$ to find $\program'$ s.t.
                $\program'\models\Examples'$.
                If we are successful in finding $\program'$, 
                then we return the program:
    \begin{algorithmic}[H]
        \Function{ForwardAllResult}{$\val$}
          \State $\val' \gets {\program}(\val)$
          \State \Return $\program'(\val')$
        \EndFunction
    \end{algorithmic}

\paragraph{Forward1 strategy:} From the program $\program$, we extract a prefix program $\program_1$ and
    we then construct a new set of examples
        $$
                \Examples' = \{ (\val'_i,\outval_i) \mid (\inval_i,\outval_i)\in\Examples, \val'_i={\program_1}(\inval)\}
        $$
        We recursively solve the PBE task $\Examples'$ to find $\program'$ s.t.
                $\program'\models\Examples'$.
                If we are successful in finding such an $\program'$, 
                then we return the program:
    \begin{algorithmic}[H]
        \Function{Forward1Result}{$\val$}
          \State $\val' \gets {\program_1}(\val)$
          \State \Return $\program'(\val')$
        \EndFunction
    \end{algorithmic}

            \paragraph{Backward1 strategy:} From the program $\program$, we extract a suffix $\program_2$.
    Note that $\program_2$ will be using some set of values computed by the prefix of $\program$.
                Without loss of generality,
    let $\val_a$ and $\val_b$ be the two values used by $\program_2$.
    We construct two new sets of examples, $\Examples_a$ and $\Examples_b$ as follows:
    for every $(\inval_i,\outval_i)\in\Examples$, if
                $\outval_i = {\program_2}((\val_a, \val_b))$,
        then
        $(\inval_i,\val_a)$ is added to $\Examples_a$
        and
        $(\inval_i,\val_b)$ is added to $\Examples_b$.
        We recursively solve the PBE tasks $\Examples_a$ and $\Examples_b$
        to find $\program_a$ and $\program_b$ s.t.
                $\program_a\models\Examples_a$ and
                $\program_b\models\Examples_b$.
        If we are successful in finding such an $\program_a$ and $\program_b$, 
                then we return the program:
    \begin{algorithmic}[H]
        \Function{Backward1Result}{$\val$}
          \State $\val_a \gets {\program_a}(\val)$
          \State $\val_b \gets {\program_b}(\val)$
          \State \Return ${\program_2}((\val_a, \val_b))$
        \EndFunction
    \end{algorithmic}
                Note that creating $\Examples_a$ and $\Examples_b$ requires executing the ``inverse'' semantics of $\program_2$. We use an LLM
                to perform that step and use the forward execution of $\program_2$ to verify the prediction of the LLM.

It is easy to see that the above approach is sound: if we ever return a program $\program$ for a given PBE task $\Examples$, then the returned program will satisfy the given examples, that is, $\program\models\Examples$.

\ignore{

The soundness of the proposed synthesis approach follows from the soundness of the inference rules
stated in Proposition~\ref{prop:soundness}.
\begin{corollary}[Soundness of Synthesis]\label{cor:soundness}
    Starting from the configuration $\synconfig{\_, \Examples}$, if the synthesis approach terminates with 
    success and returns a program $F$, then it is the case that $F\models\Examples$.
\end{corollary}

\endignore}

Our approach is recursive.
In our experiments, we restrict the recursion depth to $1$ for efficiency.
For the proposed approach to succeed, it should be the case that
(1a) either the prefix or suffix of the wrong program 
should be the prefix or suffix of some correct program, or
(1b) the wrong program should be one of the branches of a correct conditional program, and
(2) we should successfully create the subproblems, and
(3) the LLM should successfully solve the subproblems.

\section{Decomposing PBE Tasks}
\label{sec:decomposing}


A key step in our approach is the decomposition of a (wrong) program into two parts, which we have
informally called the prefix and suffix of the program.
Here we describe the details of this decomposition and the extraction of the forward1 program and
the backward1 program. 

\ignore{
We had defined sequential composition operator, $\circ$, in Section~\ref{sec:composition}.
By a (sequential) decomposition of a program $\program$, we mean breaking it into parts 
$\Programs_1$ and $\program_2$ such that the sequential composition 
$\Programs_1 \circ \program_2$ of the parts becomes equal to $\program$.
Thus, a \emph{decomposition} of a program $\program$ on inputs, say $\invar_1$ and $\invar_2$,
is a tuple $(\Programs_1,\program_2)$ where
say $\Programs_1 = \langle\program_{11},\program_{12}\rangle$, 
such that 
$$\program_2( 
    \program_{11}(\invar_1, \invar_2), 
    \program_{12}(\invar_1, \invar_2) 
      ) == \program(\invar_1, \invar_2)
$$
\endignore}

\begin{figure}[t]
\begin{center}
\centerline{\includegraphics[scale=0.30,trim=0 50 0 50,clip]{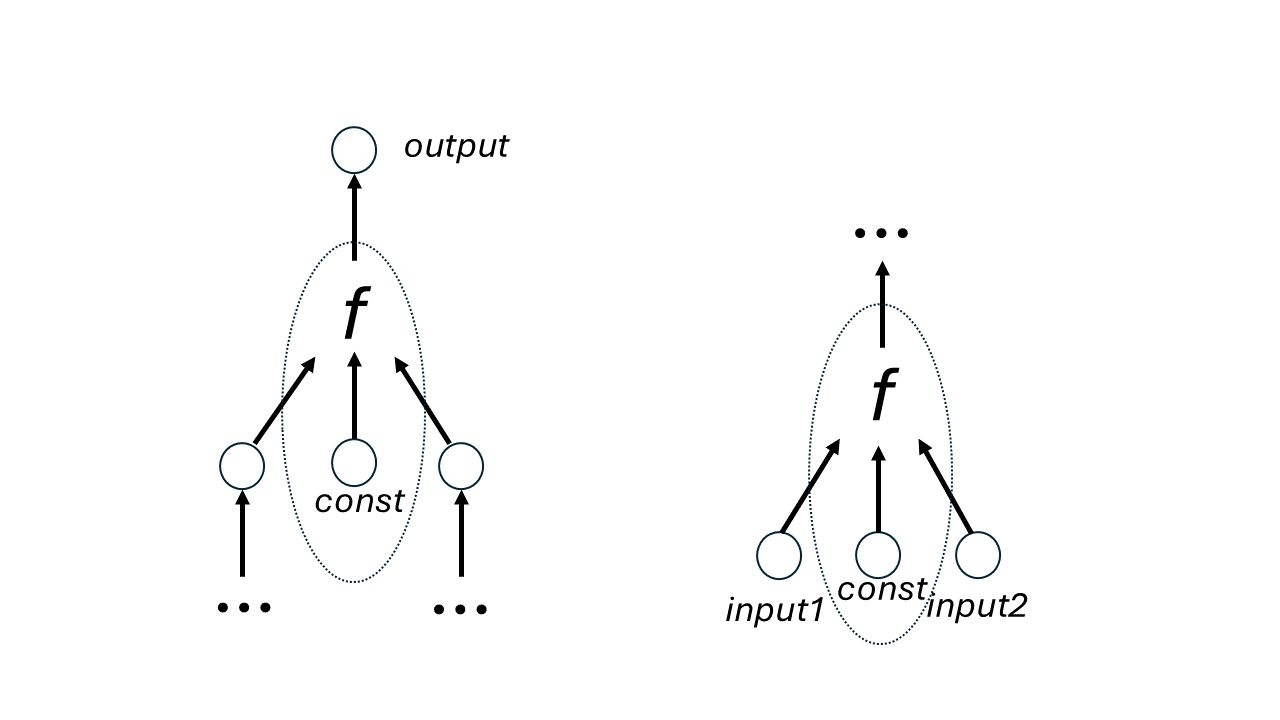}}
    \caption{\footnotesize{Decomposing a program: If we view the computation of the output from the inputs as a tree whose root is the output (left) and whose leaves are constants and inputs (right), then the backward1 program is the subtree shown within dotted oval on the left and the forward1 program is the subtree shown within the dotted oval on the right.}}
\label{fig:decomposing}
\end{center}
\end{figure}

We extract prefix and suffix for purposes of decomposition through the control flow graph
of the program. Consider a program $\program$ that computes output $\outvar$ from the inputs, say 
$\inval_1$ and $\inval_2$. As shown in Figure~\ref{fig:decomposing}, we build
a tree whose leaves are the inputs (and all constants) at the bottom (shown on right), and 
whose root is the output at the top (shown on left).
Every node in this tree is either 
a \emph{value} node or a \emph{function} node.  A function node, annotated with $f$, takes the values of all its children (below) and applies $f$ on those values to get a result that becomes the value of its parent. 

\subsection{Forward1 Decomposition}

In Forward1 decomposition, the ``prefix'' we extract from $\program$, say
$\program_1$, consists of the first step performed on the inputs in the computational tree.
As shown on the right of Figure~\ref{fig:decomposing},
$f(\invar_1, c, \invar_2)$ is (one of) the first 
computation performed on (some subset of) the inputs, $\invar_1$ and $\invar_2$,
where $c$ is a constant.
Hence, a possible \emph{forward1} program, $\program_1$, is 
$\lambda{\invar_1,\invar_2}: f(\invar_1, c, \invar_2)$.
In the Forward1 strategy, the program $\program_1$ is fixed as the first step, and we 
synthesize the rest of the program recursively. The input-output examples for the rest of the program are easily obtained
by executing $\program_1$ on the inputs from the IO examples $\Examples$.

\subsection{Backward1 Decomposition}

In Backward1 decomposition, the ``suffix'' we extract from $\program$, say
$\program_2$, consists of the last step that generates the output in the computational tree.
As shown on the left of Figure~\ref{fig:decomposing},
$f(\invar_1, c, \invar_2)$ is the last 
computation performed because it's result is the output $\outval$.
Hence, a possible \emph{backward1} program, $\program_2$, is 
$\lambda{\invar_1,\invar_2}: f(\invar_1, c, \invar_2)$.
In the Backward1 strategy, the program $\program_2$ is fixed as the last step, and we 
synthesize the remaining program recursively. We synthesize two subprograms --
one that computes $\invar_1$ and another that computes $\invar_2$.

It is challenging to find the input-output examples for the two subprograms.
It requires us to back propagate the output values through $\program_2$. 
Back propagating an output value $\outval$ means finding 
values $\val_1$ and $\val_2$ s.t.  $f(\val_1, c, \val_2) = \outval$.
Note that there could be 
multiple such $\val_1,\val_2$ that all produce the same output $\outval$. 
Ideally, we want to pick the values $\val_1$ and $\val_2$ that are {\em{most likely}} to be 
generated from the input $\inval$. 
We use a large language model (LLM) to generate the candidate values $\val_1$, $\val_2$ and then we pick
the pair that actually generates $\val$ by executing $f$ on the candidate inputs. The LLM prompt for back propagation is shown in Figure~\ref{fig:backprop-prompt}. The prompt is designed to be used with the chat completion API of LLMs, and contains a system message that describes the task and a couple of ``few-shot'' examples that provide illustration of the task. 
We ask the model to generate multiple responses at a temperature of 0.4.

\begin{figure}[tbh]
    \begin{description}
        \item[system:] {\em{You are an expert Python programmer. You are given a Python expression and some output values it computes. For every output value, your task is to provide the inputs on which the expression evaluates to that output value. For each output value, as a help, you are also given some additional values that can be used as inspiration for predicting the inputs.}}
        \item[user:] \texttt{\{'expr': 'new\_var.split(":")[0]', 
                       'outputs': [\{'output': 'foo', 'additional values': ['1234,foo:bar']\},
                                   \{'output': 'show', 'additional values': ['4356,show:full']\}]\}}
        \item[assistant:]  \texttt{[\{'output': 'foo', 'inputs': 'new\_var = "foo:bar"'\},
                                   \{'output': 'show', 'inputs': 'new\_var = "show:full"'\}]}
        \item[user:] \texttt{\{'expr': 'f"\{new\_var0\} \{new\_var1\}"',
                   'outputs': [\{'output': 'bar baz', 'additional values': ['try, baz, bar, me']\},
                               \{'output': 'joe smith', 'additional values': ['done, smith, joe, cse']\}]\}}
       \item[assistant:] \texttt{[\{'output': 'bar baz', 'inputs': 'new\_var0 = "bar"\textbackslash n new\_var1 = "baz"'\},
                  \{'output': 'joe smith', 'inputs': 'new\_var0 = "joe"\textbackslash n new\_var1 = "smith"'\}]}
       \item[user:] \texttt{\{'expr': $\langle f(new\_var0, c, new\_var1)\rangle$,
                   'outputs': [\{'output': $\langle v_1\rangle$, 'additional values': $\langle\st_1\rangle$\},
                       \{'output': $\langle v_1'\rangle$, 'additional values': $\langle\st_1'\rangle$\}]\}}
    \end{description}
    \caption{\footnotesize{Prompt for back propagating values through backward1 functions. The prompt consists of a system prompt, followed by two examples showing what the user might say and how the assistant is supposed to reply. The last message from the user is instantiated to the actual values.}}\label{fig:backprop-prompt}
\end{figure}





\subsection{Condition Learning}

In the IfThenElse strategy, the challenging step is the generation of the
condition $c$. Recall that the condition learning problem is as follows: 
given 
a program $\program_1$ that works on examples $\Examples_1$,
and
a program $\program_2$ that works on examples $\Examples_2$,
find a condition $c$ that can be used to decide whether to route the input to 
$\program_1$ or $\program_2$.


\begin{figure}[tbh]
    \begin{description}
        \item[system:] {\em{You are an expert Python programmer. Your task is to find generalized predicates or conditions that will distinguish two classes of program states. A program state is a dictionary from variable names to values.}}
        \item[system:] {\em{You should return a python function that takes a program state as input and outputs a Boolean value. It should output True if the input state is one of states in Class1 or it shares some commonality with all the sample states in Class1. It should return False if the input state is in Class2 or it shares some commonality with all sample states in Class2.}}
        \item[user:] \texttt{class1\_sample\_states = [\{...\}, \{...\},...]}
        \item[user:] \texttt{class2\_sample\_states = [\{...\}, \{...\},...]}
    \end{description}
    \caption{\footnotesize{Prompt for generating conditions for use in if-then-else programs to enable parallel composition of two programs to yield a correct final program. Any generated candidate programs can be executed to check for their correctness.}}\label{fig:condition-prompt}
\end{figure}

We use LLM again to generate the condition. The prompt for condition learning is shown
in Figure~\ref{fig:condition-prompt}. We observe that the prompt instructs the model to generate
a Python function. There is emphasis on learning generalized predicates since the LLM has a tendency
to learn conditions that are overfitted to the values provided in Class1 and Class2. The word 
``sample'' also helps the model realize that values provided in the classes are only samples, and so
it should generate conditions based on the common patterns seen across the samples.
We finally note that any Boolean valued Python program returned by the model can be checked for
its correctness by just executing it on the elements in Class1 and Class2.

\subsection{Trigger Conditions}

We described four synthesis strategies, namely
\textsc{ForwardAll}, 
\textsc{Forward1}, 
\textsc{Backward1}, and
\textsc{IfThenElse}.
However, which one to use and which order to try them was left unspecified.

The generic precondition that guards the application of the \textsc{Forward1}, \textsc{ForwardAll},
and \textsc{Backward1} strategy is that
there should exist a program $F$ that is not a solution for $\Examples$.
Additionally for \textsc{Forward1}, we need to have a nonempty set of forward1 programs,
$\Programs_1$, extracted from $F$, and moreover, these programs should generate non-$\bot$ values when executed on
the inputs in the examples $\Examples$. 
If these checks are violated, then \textsc{Forward1} is not applicable.
Analogous preconditions can be easily written for the \textsc{Backward1} strategy
and the \textsc{ForwardAll} strategy.
We focus on evaluating the strategies whenever they are applicable, and do not focus on 
trigger conditions for the strategies in this paper.

\ignore{
Beyond the generic preconditions, we can also have domain-specific heuristics to 
quickly determine the synthesizability of a PBE (sub)task.
Our experiments are performed over the domain of string manipulation tasks.
For problems in this domain, we observe that for every synthesizable PBE task, the output is made
up of pieces where each piece either comes from the input, or it is a constant string that also occurs
in other outputs, or it comes from some standard list of special words.
For example, if we consider the example where an input \texttt{John Smith 5/2024} is transformed to
an output \texttt{Mr. Smith May 2024}, we find that \texttt{Smith} in the output comes from the input,
\texttt{Mr.} comes from a constant string that will also occur in the outputs of other examples, and
\texttt{May} comes from a collection of special words that contain names of all the months, weekdays, etc.
Thus, for this domain, the following is a quick check for synthesizability of a PBE task:
\\
(1) for every example input-output example
$(\st_i,\val_i)$ in the PBE task $\Examples$, 
we first tokenize the output string $\val_i$ (using any off-the-shelf tokenizer)
to get, say, $t_{i1}\ldots t_{ik}$
\\
(2) we declare the PBE task as not synthesizable if there exists a token $t_{ij}$ in the 
tokenization of some $\val_i$ such that: 
 (a) the string $t_{ij}$ does not occur (as a substring) in any value in the corresponding input $\st_i$, and
 (b) the string $t_{ij}$ does not occur (as a substring) in any \emph{other} output $v_l$ for $l\neq i$, and
 (c) the string $t_{ij}$ does not occur (as a substring) in a curated list of special words.
\\
We can use the above synthesizability checker to avoid pursuing a synthesis strategy that is unlikely to lead to
success.

\endignore}



\section{Evaluation}
\def\flashfill{\texttt{FlashFill}}
\def\tformula{\texttt{Conditionals}}

We evaluated the four strategies on their ability to synthesize programs from
input-output examples where more classic approaches failed.

\subsection{Benchmarks}
\label{sec:benchmarks}

The Playgol dataset~\cite{playgol} consists of a collections 327 real-word string transformation tasks, 
which has some lineage to the datasets from~\cite{lin2014,gulwani2011automating}. Each task is given
as a set of input-output examples. There are around 10 examples in each task. Each example is a pair
consisting of a list of input strings and a single output string. Most tasks have exactly 1 input.

To create a PBE task, we need to partition the ~$10$ examples into train and test sets.
The train set $\Examples$ defines a PBE task and is used to synthesize a program.
The generated program is evaluated \emph{both} on the train set $\Examples$ and the examples
in the test set. This evaluation guarantees that the learned programs are not over specialized
to the examples in the train set $\Examples$ and can work on other similar inputs.

We pick $18$ different ways to partition the examples into train and test sets to get a starting set of $327*18$ PBE tasks.
Next, we filter this set and remove the ``easy'' PBE tasks. 
First, we remove the PBE tasks that are solved by a single call to an LLM with a suitable prompt.
Next, we run a self-reflection approach that tries to redo the task up to $4$ additional times using LLM calls. Each LLM call tries to fix the wrong program generated in the previous call, given the error generated by the execution of the wrong program.
The error could be an exception thrown, or one of the examples that was incorrectly solved.
This approach is a powerful and popular approach to boost the precision of LLMs~\cite{reflexion,chen2023teaching,llm-cegis}. 
If it was able to find the correct program (that works on all train and test examples) in any of the iterations,
then that PBE task is also deemed ``easy'' and removed.
Since the self-reflection procedure generates and repairs code in some target language, the tasks removed will depend
on the target language used. 
When we use Python as the target, after removing the easy tasks as described above, we were left with 665 PBE tasks, which form our \emph{\sysname-playgol-py benchmark} set.  
When we use Excel formulas as the target, after the same process, we were left with 1134 PBE tasks, which form our \emph{\sysname-playgol-xl benchmark} set.  


\ignore{ 
\subsection{Implementation Details}
\label{sec:implementation}
A critical and nontrivial component in the implementation of our approach is the computation of the forward1 and backward1
programs from a given program.  We described the process conceptually in Section~\ref{sec:decomposing}.

In the concrete implementation,
for Python, we used the inbuilt python module \texttt{ast} for parsing a
Python program. Extracting the forward1 and backward1 programs from the parsed ast tree is still nontrivial.
This is because the abstract syntax tree of a Python program can be fairly complex, representing function 
declarations, assignments, and nested expressions. One has to build the dataflow graph to map the AST
to the computation graph described in Section~\ref{sec:decomposing}.
For example, if the input is $A$ and the Python code computes \texttt{A.split()[-2:]}, then 
one possible forward1 program is \texttt{A.split()}.
Similarly, if the Python program contains \texttt{res = f(a, b); return res}, then backward1 program
will be \texttt{f(a,b)}. 

One significant challenge comes from programs that have control-flow split and merge; for example, when 
there are if-then-else and loops in the program. For such programs, in our
implementation, whenever there is a choice, we just pick one dataflow path from the inputs to the output.
We do note here that there is no notion of a correct or wrong forward1 or backward1 program. 
Some choices for a backward1 or forward1 program can cause synthesis to fail, but it can not cause
synthesis to be unsound.

The ability to execute code is required to create the subproblems.
For Python, we used the inbuilt functions
\texttt{eval} and \texttt{exec} to execute Python code. 

For Excel, we used an open-source pypi module for parsing and executing Excel formulas when we had
to compute and execute forward1 programs to create subproblems, 
as well as when we had 
to compute and execute backward1 programs to create subproblems.
For Excel formulas, the parse tree was easier to work with since AST represented just one expression
the computed the output from the inputs. Since we did not want to deal with datatypes other than strings,
we picked forward1 programs that returned string values, but that were minimal (smallest) such programs.
Unfortunately, the open-source parser and execution engine was incomplete, which means that we very
often just failed to compute the forward1 or backward1 programs, or the resulting subproblems.
This has an impact on the results as we will see below.

\endignore}

\subsection{\sysname\ Solves Challenging PBE Problems}
\label{sec:rq1}

The first research question asks whether our failure-guided compositional program synthesis approach can solve
challenging PBE problems. Since the \sysname-playgol-py benchmark set was constructed by removing the easy PBE tasks, we 
answer this research question by evaluating the performance of \sysname\ on this benchmark set.

\begin{figure}[t]
    \begin{tabular}{cc}
        \small{
        \begin{minipage}{0.25\textwidth}
            \begin{tabular}{|l|c|c|c|}
                \toprule
                  & \#LLM & 
                  \multicolumn{2}{c|}{Solved}
                  \\ & Calls & \# & \%
                \\ \midrule\midrule
                \textsc{FwdAll} & 2 & 115 & 17.4\%
                \\ \midrule
                \textsc{Fwd1} & 2 & 86 & 12.9\%
                \\ \midrule
                \textsc{Bwd1} & 3-4 & 82 & 12.3\%
                \\ \midrule
                Any one & $<6$ & 210 & 31.6\%
                \\ \bottomrule
            \end{tabular}
        \end{minipage}
    }
    &
        \begin{minipage}{0.2\textwidth}
\begin{center}
\centerline{\includegraphics[scale=0.45,trim=0 0 0 0,clip]{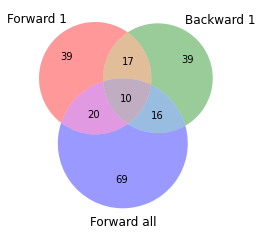}}
\end{center}
        \end{minipage}
    \end{tabular}
\caption{\footnotesize{The number of benchmarks solved in Python by each technique and their union -- out of the 665 \sysname-playgol-py {\em{hard}} benchmarks. 
The Venn diagram showing the same data, which also highlights how many benchmarks are solved exclusively by one or two of the three approaches. Each technique is able to solve some benchmarks that no other technique solved.}}
\label{fig:venn}
\end{figure}

The target language for synthesis was Python for these experiments. The model used here was Gpt-4o.
Figure~\ref{fig:venn} shows the performance of \textsc{ForwardAll}, \textsc{Forward1} and \textsc{Backward1} on the filtered
665 \sysname-playgol-py benchmark set.  For these experiments, starting with the wrong program generated by the LLM,
we use each strategy exactly once to synthesize the correct program. The subproblems generated by each strategy
are directly solved using one LLM call. If the LLM fails to solve a subproblem, then we just fail; that is, we \emph{do not} 
use any of the techniques iteratively to again repair the wrong subprogram. The reason for performing just one iteration
is to limit the computational cost of performing experiments. 
The Columns ``\#LLM calls'' gives the number of LLM calls used per benchmark. There is 1 call used for generating the initial
(wrong) program, and then one more to solve the subproblem in \textsc{Forward1} and \textsc{ForwardAll}. 
For \textsc{Backward1}, we need 1 additional call to backpropagate output values, and additional 1-2 LLM calls depending on how many subproblems
are created (we get at most 2 subproblems). In the ``Any one'' row, we stop at first success, so the total number of LLM calls per benchmark is upper bounded
by $6$, but the exact number will vary between $2$-$6$.

Even with just one application of problem decomposition, we are able to solve $210$ benchmarks out of the $665$, which is
31.6\%, or about a third of the benchmarks.
The table in Figure~\ref{fig:venn} provides the raw numbers of benchmarks solved by each technique separately and the
Venn diagram on the right of the figure shows the overlap between the benchmarks solved by the three techniques.
Clearly, each technique was able to solve some benchmarks solved by no other technique. This shows the unique value
brought by each technique, and it also provides some validation for the intuitions behind the design of these techniques.

\subsection{If-Then-Else Strategy on Conditional Benchmarks}
\label{sec:rq2}

The \sysname-playgol-py benchmarks contain string processing tasks that are not ideal candidates for learning conditional
programs. Therefore, we picked two other datasets of string processing tasks: \flashfill\ and \tformula.
On the \flashfill\ benchmarks~\cite{cambronero2023flashfill++}, we performed the same preprocessing step that we did
on Playgol to get the \sysname-\flashfill-py benchmarks (where self-reflection failed to find a correct Python program). 
The \tformula\ benchmark is designed specifically for testing the ability to synthesize conditional programs.
We did not multiply this benchmark set, but we did remove the instances where self-reflection worked to get
the filtered \sysname-\tformula-py benchmarks.

\begin{table}[t]
    \small{
    \begin{tabular}{|l|c||c|c|c||c|c||c|}
        \toprule
        Dataset \sysname- & Size & Ph0 & Ph1 & Ph2 & Ph3 & \% 
        \\ \midrule \midrule
        -playgol-py & 665 & 637 &  221 & 43 & 25 & 11.3 
        \\ \midrule
        -\flashfill-py & 658 & 597 & 422 & 134 & 55 & 13.0 
        \\ \midrule
        -\tformula-py & 29 & 26 & 20 & 5 & 4 & 20.0 
        \\ \bottomrule
    \end{tabular}
}
    \caption{\footnotesize{Performance of the If-Then-Else strategy on different benchmarks in Python: The first numerical column is the number of benchmarks and each subsequent column shows the number of benchmarks that remain after each significant phase of the strategy. Column ``Ph3''  reports the successful benchmarks. The success percentages, shown in the last column, are calculated over ``Ph1''.}}
    \label{table:ite}
\end{table}

The target language for synthesis was Python for these experiments. The model used here was Gpt-4o. 
Table~\ref{table:ite} shows the results from using the If-Then-Else strategy on the benchmarks. We use exactly 3
LLM calls on each benchmark: the first for generating the first branch $F_1$, the second for the second branch $F_2$, and the
third for generating the condition $c$. 
If $F_1$ execution terminates successfully on some examples, then that benchmark is included in \texttt{Ph0}.
For each benchmark in \texttt{Ph0}, 
if $F_1$ correctly solves some, but not all, examples, then that benchmark is included in \texttt{Ph1}.
For each benchmark in \texttt{Ph1}, 
if $F_2$ correctly solves all the remaining examples (left unsolved by $F_1$), then that benchmark is included in \texttt{Ph2}.
Finally,
for each benchmark in \texttt{Ph2}, 
if the generated condition $c$ accurately separates the examples on which $F_1$ and $F_2$ work correctly, then that benchmark is included in \texttt{Ph3},
which are the successful benchmarks.
The condition $c$ is generated by prompting an LLM with the prompt shown in Figure~\ref{fig:condition-prompt}. 

We observe that the If-Then-Else strategy has a lower success rate than the other three strategies on the
same \sysname-playgol-py  benchmarks. This is because those benchmark sets do not contain enough
tasks that require conditional programs. On the benchmark set that contains tasks that require conditional
programs, namely \sysname-\tformula-py, the success rate for If-Then-Else strategy improves as expected, thus validating the strategy. 

\subsection{\sysname\ Adapts to a Different Target}
\label{sec:rq3}

We repeat the experiment reported in Section~\ref{sec:rq1}, but now change the target language to Excel formulas. 
We make minimal changes in the prompts to indicate that the target language is Excel (and not Python). Some challenges due to non-availability of a reliable parser and execution engine for Excel clouds our results.

\begin{figure}[t]
    \begin{tabular}{cc}
        \small{
        \begin{minipage}{0.25\textwidth}
    \begin{tabular}{|l|c|c|c|}
        \toprule
         & Fwd1 & Bwd1 & FAll 
        \\ \midrule
        Ph1 & 1058 & 1089 & 1134 \\
        Ph2 & 927 & 485 & 1020 \\
        Ph3 & 81 & 128 & 93 \\
        \% & 7.7 & 11.8 & 8.2 \\
        \bottomrule
    \end{tabular}
        \end{minipage}
    }
        &
        \begin{minipage}{0.2\textwidth}
\begin{center}
\centerline{\includegraphics[scale=0.4,trim=0 0 0 0,clip]{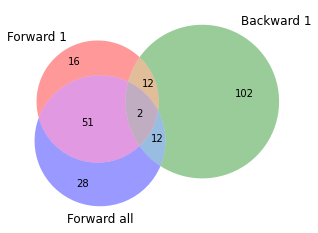}}
\end{center}
        \end{minipage}
    \end{tabular}
    \caption{\footnotesize{Performance results for Excel on 1134 benchmarks from filtered \sysname-playgol-xl dataset. The rest of the setup is the same as for Figure~\ref{fig:venn}. The success percentages for three strategies, \textsc{Forward1}, \textsc{Backward1} and \textsc{ForwardAll}, are shown in brackets alongside the total number
    of benchmarks that were solved by that strategy. Success percent is calculated over ``Ph1'' values.
    Venn diagram of the same result. The three techniques combined
    succeed on 223 benchmarks out of 1134, which is $19.7\%$ success rate.}}
    \label{fig:venn-xl}
\end{figure}

Figure~\ref{fig:venn-xl} shows the results of using \sysname\ for synthesizing Excel formulas from examples on the
\sysname-playgol-xl dataset. 
For each of the three strategies, Phase1 (or ``Ph1'') refers to extraction of the part of the program that is preserved, Phase2 (or ``Ph2'') refers to creation of the subproblems, and Phase3 (or ``Ph3'') refers to solving the subproblems. 
We report number of benchmarks that reached that phase.
Thus, the number under ``Ph3'' are the final success numbers for that strategy. 
For \textsc{forwardAll}, to create the subproblem, we need the program $F$ generated by the first LLM call to execute successfully on the inputs in the train set. 
Out of the $1134$ benchmarks, this happened for $1020$ benchmarks, and out of those 
\textsc{forwardAll} solved $93$ benchmarks.
For \textsc{forward1}, in Phase1 we have to parse the program $F$ and extract the prefix $F_1$,
which was possible for $1058$ benchmarks. Thereafter, we had to execute the forward1
subprograms $F_1$ to get the subproblems, which succeeded for $927$ benchmarks. Finally, we had to 
solve the subproblems, which happened  successfully for only $81$ benchmarks. 
Similarly, for \textsc{backward1} strategy, we were able to parse $F$ and get the backward1 function $F_2$ for $1089$ benchmarks, but we were able to use the LLM to generate subproblems for only
$485$ benchmarks, and we could solve those subproblems successfully on $128$ benchmarks.
Figure~\ref{fig:venn-xl} also shows the Venn diagram for the three \sysname\ synthesis strategies.
Compared to Python, we observe a larger overlap between the sets where
\textsc{Forward1} and \textsc{ForwardAll} succeed. Nevertheless, we see an overall success rate
of $19.7\%$ when the three techniques are combined.

\ignore{
\begin{figure}
\centerline{\includegraphics[scale=0.45,trim=0 0 0 0,clip]{venn-xl}}
    \caption{\footnotesize{Venn diagram showing the performance of 
    \textsc{Forward1},
    \textsc{ForwardAll},
    and
    \textsc{Backward1} on the \sysname-playgol-xl dataset. The three techniques combined
    succeed on 223 benchmarks out of 1134, which is $19.7\%$ success rate.}}\label{fig:venn-xl}
\end{figure}

\endignore}

The evaluation numbers establish that even for a different target language, Excel formulas, 
the \sysname\ approach is able to solve benchmarks that are left unsolved by 5 iterations of
self-reflection. 
Apart from less familiarity of LLM with Excel, the
low numbers for Excel was also  due to some technical challenges coming from working with an incomplete parser and execution engine.

\ignore{ 

\subsection{Effectiveness of Synthesizability Checker}
We now evaluate the effectiveness of the synthesizability checker in helping us avoid trying
a synthesis strategy.
We perform the evaluation by looking at all the PBE subtasks generated by 
\textsc{Forward1} and \textsc{ForwardAll}.
The \textsc{Backward1} strategy actually fails to generate subtasks frequently, but when
it generates subtasks, there is a high chance that they are solvable. For example,
in the \sysname-playgol-py dataset, \textsc{Backward1} generates subtasks for only
$112$ benchmarks, and out of those, it succeeds on $82$.
On the other hand, \textsc{Forward1} generates subtasks for all the $665$ benchmarks 
(because the incorrect $F$ parses and generates forward1 programs that execute successfully in all cases) and
\textsc{ForwardAll} generates subtasks for $638$ benchmarks (because the incorrect $F$ executes successfully
on only those many).

\begin{table}
    \small{
    \begin{tabular}{|l|c|c||c|c|}
        \toprule
        & \multicolumn{2}{|c||}{\textsc{Forward1}} & \multicolumn{2}{|c|}{\textsc{ForwardAll}}
        \\
        & Synthesizable & Not Synthesizable  & Synthesizable & Not Synthesizable
        \\ \midrule
        \# subtasks that succeed & 86 & 0 & 81 & 34
        \\
        \# subtasks that fail & 377 & 202  & 264 & 259
        \\ \bottomrule
    \end{tabular}
}
    \caption{\footnotesize{Efficacy of the synthesizability checker: If we had a perfect predictor of synthesizability,
    all off-diagonal terms in the two $2\times 2$ matrices would be zero. 
    The off-diagonal numbers indicate where the synthesizability checker makes an incorrect prediction.}}
    \label{table:checker}
\end{table}

We use the synthesizability checker to mark each subtask as being synthesizable or not, and compare it with
the actual finding of whether the subtask succeeded or not. 
Table~\ref{table:checker} shows the results. 
The first $2\times 2$ matrix contains the numbers for subtasks created by \textsc{Forward1} and
the second $2\times 2$ matrix contains the numbers for subtasks created by \textsc{ForwardAll}.
The off-diagonal numbers show disagreement between the synthesizability checker and the actual 
behavior observed in our experiments. Only the number in the top-right of the $2\times 2$ matrix is 
worrisome since for those cases, the checker would mark them as ``not synthesizable'', but in reality
we would succeed. For \textsc{Forward1}, that number is zero. For \textsc{ForwardAll}, we have 34 
benchmarks that we would have (incorrectly) failed on if we followed the checker's suggestion.
On manual inspection, it turned out that almost all these cases had synthesized programs that were
unnatural that used conditionals to overfit to particular examples (even though the passed the
held out test examples). 
The savings from using the synthesizability checker are enormous: it helps avoid exploring
almost 50\% of the cases that lead to failures.

\endignore}

\section{Related Work}
\paragraph{Programming by Example}

Learning to write programs from demonstrations has been a popular research area for a long time \cite{cypher1993watch}.
Two popular approaches are \emph{bottom-up}, or forward, synthesis and \emph{top-down}, or backward, synthesis \cite{gulwani2017program}.
FlashFill \cite{gulwani2011automating}---probably the most famous PBE system---uses backward synthesis in combinations with \emph{witness functions}, which reverse the semantics of operators in the grammar, to break the problem into smaller sub-problems until each sub-problem can be solved.
FlashFill++ \cite{cambronero2023flashfill++} and other recent approaches \cite{lee2021combining} combine top-down with a bottom-up component that scans the input for patterns that are likely leaf nodes, for examples, extracting numbers or date components from input strings.
In this paper, we take the ideas from forward and backward synthesis, and mesh them into an LLM-based approach for synthesis. The LLM-generated candidate program guides the choice of subprogram (in forward synthesis) and the cap (in backward synthesis), and the subproblems are again handled by LLMs.

\paragraph{Programming by Example with Neural Networks}

Both forward and backward synthesis need to explore vast search spaces.
Statistical approaches leverage data to learn likely combinations of operators---often conditioned on the examples---to drive the search through this space.
For forward synthesis, these approaches learn to predict both the production rules and its inputs \cite{ellis2019write,ellis2021dreamcoder}.
For backward synthesis,  neural networks can be used to assign probabilities to production rules in the grammar for a guided search \cite{kalyan2018neural,parisotto2022neuro}.
Other approaches do not perform a search and have the network directly predict tokens \cite{robustfill}.
In this paper, the LLM does most of the work, and symbolic reasoning supports it by decomposing problems.

\paragraph{Programming by Example with Large Language Models}

Large language models are large, auto-regressive models trained on huge corpora of text \cite{brown2020language}.
By asking them to generate hypotheses in natural language and then translating these to code, they become better at programming by example \cite{wanghypothesis}.
By fine-tuning them, large language models surpass or approximate most symbolic systems on their domains \cite{li2024programming} \emph{if} enough programs are sampled from the fine-tuned language model.
Additionally, they can be used to augment symbolic programming-by-example systems with semantic capabilities \cite{verbruggen2021semantic}.
In this paper, we do not fine-tune or train models, but use popular LLMs in tight integration with symbolic reasoners, which is
unlike any existing work.

\ignore{ 
\paragraph{Counter-example Guided Repair using LLMs}
Large language models have been used to repair programs; see for example~\cite{chen2023teaching,llm-cegis}
among many others.
Counterexample-guided generation is also an instance of the self-debugging approach
outlined in~\cite{chen2023teaching}. 
The idea in these approaches is to have the LLM do a task, but if the artifact produced by the
LLM fails to satisfy some checkable constraints, such as whether the program parses or executes without
throwing an exception, then the LLM is invoked again. In the second call, the LLM is provided a
reason for why the previously generated artifact was not accepted.
This process can be repeated until the LLM generates the correct output or a limit on 
the number of LLM calls is reached. 
We call this approach counter-example-guided synthesis (CEGIS), and we use it as a baseline. 
A key difference with CEGIS is that each subsequent LLM call in our case is solving a different
and potentially simpler task (and not the same task).


\endignore}

\section{Conclusion}

We presented \sysname\ -- an approach for
LLM-based by-example program synthesis. 
\sysname\ consists of four main program synthesis strategies. We showed that
these strategies together can solve around 30\% of the benchmarks that are not
solved by self reflection. There is a lot of room for customization by picking different tactics to apply the
strategies. 
There are also choices related to deciding exactly what program parts are extracted as
forward1 and backward1 programs. We presented one particular way, but other options
may perform better, especially for domains other than string transformation.
\sysname\ can also be viewed as a generic approach for improving generation quality of LLMs.
We have worked out the details for \sysname\ in the context of PBE. It will be interesting
to explore the ideas underlying \sysname\ on other classes of tasks. 
Implementing the \sysname\ framework in an agentic  framework~\cite{autogen} is also an exciting 
avenue for future work.

\bibliographystyle{named}
\bibliography{ijcai25,bib-flashgpt}

\end{document}